\documentclass[preprint,showpacs,preprintnumbers,amsmath,amssymb]{revtex4}

\usepackage{graphicx}
\usepackage{dcolumn}
\usepackage{bm}
\usepackage{subfigure}
\usepackage{color}
\usepackage{mathrsfs}
\newcommand{\Cr}{CrNb$_3$S$_6$}
\usepackage{ulem}

\begin{document}

\title{Fan-type spin structure in uni-axial chiral magnets 
}
\author{M.~Shinozaki$^1$}
\author{S.~Hoshino$^2$}
\author{Y.~Masaki$^3$}
\author{A.~N.~Bogdanov$^{1,4}$}
\author{A.~O.~Leonov$^{5,6}$}
\author{J.~Kishine$^{5,7}$}
\author{Y.~Kato$^{1,3}$}
\affiliation{$^1$Department of Basic Science, The University of Tokyo, Meguro-ku, Tokyo 153-8902, Japan}
\affiliation{$^2$Center for Emergent Matter Science, RIKEN, Wako-shi, Saitama, Tokyo 351-0198, Japan}%
\affiliation{$^3$Department of Physics, The University of Tokyo, Bunkyo-ku, Tokyo 113-0033, Japan}
\affiliation{$^4$IFW Dresden, Postfach 270016, D-01171 Dresden, Germany}
\affiliation{$^5$Center for Chiral Science, Hiroshima University, Hiroshima university, Higashi-Hiroshima, Hiroshima, 739-8526, Japan}
\affiliation{$^6$Department of Chemistry, Hiroshima university, Higashi-Hiroshima, Hiroshima, 739-8526, Japan}
\affiliation{$^7$The Open University of Japan, Mihama-ku, Chiba 261-8586, Japan}%
\date{\today}
%
\begin{abstract}
We investigate the spin structure of a uni-axial chiral magnet near the transition temperatures in low fields perpendicular to the helical axis.
We find a fan-type modulation structure where the clockwise and counterclockwise windings appear alternatively along the propagation direction of the modulation structure. This structure is often realized in a Yoshimori-type (non-chiral) helimagnet but it is rarely realized in a chiral helimagnet. 
To discuss underlying physics of this structure, we reconsider the phase diagram (phase boundary and crossover lines) through the free energy and asymptotic behaviors of isolated solitons.  
The fan structure appears slightly below the phase boundary of the continuous transition of instability-type. In this region, there are no solutions containing any types of isolated solitons to the mean field equations.
\end{abstract}
\maketitle

\section{\label{sec:level1}Introduction}
Helical spin structures have been found in a number of magnetic compounds~\cite{Izyumov}. Those are categorized into two parts according to the existence/absence of a space-inversion symmetry; 
one is the symmetric helimagnet resulting from the competition between the nearest-neighbor interaction and the antiferromagnetic next-nearest-neighbor interaction. Such a helimagnet is called Yoshimori-type ($J_1$-$J_2$ type) or non-chiral helimagnet, in which the left- and right-handed helimagnets are degenerated~\cite{Yoshimori}. 
Under an external field perpendicular to the propagation direction (helical axis) of the modulation, it forms a fan structure, in which the clockwise and counter-clockwise windings appear alternately with a constant modulus of the local magnetization~\cite{Izyumov}. 
The other kind of helimagnet is the chiral helimagnet~\cite{Dzyaloshinskii1,Dzyaloshinskii2,Dzyaloshinskii3}, which has the modulated magnetic structure induced by the competition between the Dzyaloshinskii-Moriya (DM) interaction and exchange interaction. Those structures contain the magnetic vortices (skyrmion)~\cite{Bogdanov89,Bogdanov94} as well as one-dimensional helicoids~\cite{Dzyaloshinskii1,Dzyaloshinskii2,Dzyaloshinskii3}. Those modulation structures have fixed the winding direction along the propagating direction.  In this paper, we focus on the one-dimensional modulation of the magnetic structure in the uni-axial chiral helimagnet. 

When the external field is applied to the chiral helimagnet perpendicularly to the helical axis, a modulation structure called a chiral soliton lattice (CSL) is stabilized at low temperatures~\cite{Dzyaloshinskii3,Kishine_S_T}. 
Dzyaloshinskii firstly investigated the behavior of the physical properties of the chiral helimagnets using a micromagnetic model and elucidated the critical behavior near the phase boundary~\cite{Dzyaloshinskii1, Dzyaloshinskii2, Dzyaloshinskii3}. Realization of the modulation structures in \Cr\ \ was firstly discussed by Moriya and Miyadai~\cite{Moriya}, and Miyadai et al.~\cite{Miyadai}. After thirty-years, Togawa et al. observed the modulation structure through the Lorentz transmission electron microscope experiments~\cite{Togawa1}. Stimulated by this experiment~\cite{Togawa1}, the modulation structure in the uni-axial chiral magnet becomes one of the growing fields in the chiral magnetism (see recent review articles~\cite{Kishine_Ovchinnikov,Togawa_symmetry}). 

Dzyaloshinskii in his seminal paper~\cite{Dzyaloshinskii3} argued the property of continuous phase transition between the modulation phase (i.e. CSL) and spatially uniform phase (disordered phase); the specific heat and susceptibility diverge when the phase boundary approached from the ordered phase while they do not diverge when the phase boundary approached from the spatially uniform (disordered) phase. 

This property (that criticality emerges only in one side of the phase boundary) is peculiar, compared to the conventional continuous transition, where the critical properties are the same in both sides of the phase boundary in the phase diagram. Subsequently, de Gennes~\cite{deGennes} named the former (, which Dzyaloshinskii discussed in Ref.~\cite{Dzyaloshinskii3}) the nucleation-type and the latter (i.e. conventional continuous phase transition) instability-type. 

In Ref.~\cite{Dzyaloshinskii3}, Dzyaloshinskii assumed that the modulus of the local magnetization is constant in space and considered only the spatial variation of the direction of the local magnetization. This treatment (constant-modulus approximation) is valid in a low temperature region far from the transition temperatures at low fields. In higher temperatures, we have to use the model that takes account of the spatial-, temperature-, and field-dependences of the modulus of the local magnetization (soft-modulus effect). As an earlier study on the effect of the soft modulus in magnets, we refer to the study by Bulaevskii and Ginzburg~\cite{Bulaevskii}, where they showed that the domain wall structure in a ferromagnet can be different near the transition temperature from that in low temperatures. 
Another example is the study of Schaub and Mukamel~\cite{Schaub}, where they took into account the variation of the modulus of the order parameter in a Ginzburg-Landau model that exhibits an incommensurate phase. They revealed that there are two critical points: a tri-critical point which separates the instability-type and the first-order segments and a multicritical point which separates the nucleation-type and the first-order segments. 
As the third example of study on the soft modulus, we refer to the study in a cubic chiral helimagnet~\cite{Leonov2010,Wilhelm1}, where the soft-modulus effect leads to the rich phases including the half-skyrmion lattice near the transition temperatures in low fields~\cite{Leonov2010,Wilhelm1}. 

A recent experiment on \Cr\ \ implies a critical point at a finite magnetic field~\cite{Tsuruta2}. For the uni-axial chiral helimagnet, the continuum theory of a variational mean field~\cite{Laliena1, Laliena2, Laliena3} and a mean-field (MF) theory~\cite{Shinozaki} in the lattice model have revealed a signature of the first-order transition. A Monte Carlo study~\cite{Nishikawa} found a critical point at a finite magnetic field. 

The above earlier works imply that the property of the phase transition and the spatial pattern of the magnetic structure is affected by the soft-modulus effect. We address the soft-modulus effect on the magnetic structure in the uni-axial chiral helimagnet in this paper. We will show that the magnetic structure near the transition temperatures in low fields is different from that expected as the chiral soliton lattice. The structure we find is similar to a kind of fan structure, which is common in non-chiral helimagnets but rare in chiral magnets. We attribute the stability of this fan-like structure to the reduction of the modulus of the local magnetization near the transition temperatures in low fields. Our argument is consistent with the underlying crossover lines.

In the next section, we explain the model and method we use. 
In Sect.~\ref{sec:Phase diagram}, we present the results on the phase diagram that contains the phase boundary and various crossover lines. 
We present the spin structure, which we call chiral-fan structure, in Sect.~\ref{sec:Chiral-fan}. In Sect.~\ref{sec:Conclusion}, we summarize the present study. 

\section{\label{sec:Model and method}Model and method}
\begin{figure}[tb]
	\begin{center}
		\includegraphics[width=4.5cm,clip]{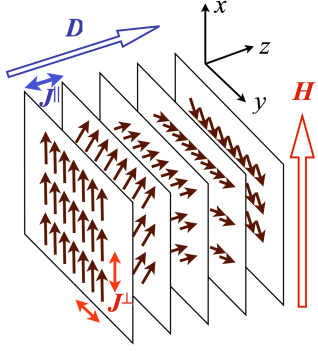}
	\end{center}
	\caption{(Color online)  An illustration for three-dimensional chiral helimagnet. }
	\label{Figure}
\end{figure}

We consider the chiral helimagnet using a lattice model defined on a cubic lattice which is schematically shown in Figure~\ref{Figure}.  
We label each site by a dimensionless vector  $\bm{i}={\bm i}_\bot +\ell \hat{\bm{z}}$, with an integer $\ell$ specifying a layer, and a two-dimensional (2d) vector $\bm i_\bot = i_x\hat{\bm x} +i_y\hat{\bm y}$ with integers $i_x, i_y$ specifying a site in the layer. 
The spin Hamiltonian we consider is 
\begin{align}
	\mathcal{H} = 
		&-J^{\bot}\sum_{\bm{i}}\bm{S}_{\bm{i}}\cdot(\bm{S}_{\bm{i}+\hat{\bm x}} + \bm{S}_{\bm{i}+\hat{\bm y}})
	\notag\\
		&-J^{\parallel}
\sum_{\bm{i}}\bm{S}_{\bm{i}} \cdot \bm{S}_{\bm{i}+\hat{\bm{z}}}
		- D \hat{\bm{z}}\cdot\sum_{\bm{i}}(\bm{S}_{\bm{i}}\times\bm{S}_{\bm{i}+\hat{\bm{z}}})
	\notag\\
		&- H \sum_{\bm{i}} \bm{S}_{\bm{i}} \cdot \hat{\bm{x}}, 
		\label{eq:hamiltonian}
\end{align}
where $\bm S_{\bm i}$ denotes a classical Heisenberg spin with the magnitude $S$ at site $\bm i$. The DM interaction vector $D\hat{\bm{z}}$ is parallel to the $z$-axis, and the ferromagnetic exchange interactions $J^\parallel$ and $J^\bot$ act respectively on the nearest neighboring pairs of spins connected by the bonds along the helical axis and inside the 2d-layer. 
The external field $H$ is applied to the $x$-axis to realize the chiral soliton lattice. 

Following the previous paper~\cite{Shinozaki}, we apply the MF method, where the system is described by a single-site spin Hamiltonian written as 
\begin{align}
	{\cal H}^{\rm MF} = -\sum_{\ell}{\bm H}_{\ell}^{\rm eff} \cdot \sum_{\bm {i}_\bot} {\bm S}_{\bm{i}_\bot+\ell\hat{\bm{z}}} + N_{2{\rm d}}\sum_{\ell} C_{\ell},
\end{align} 
with the effective field at the $\ell$-th layer 
\begin{align}
	{\bm H}^{\rm eff}_{\ell} = 
	& J^\parallel (\bm M_{\ell+1} + \bm M_{\ell-1}) + 4J^{\bot} \bm M_{\ell}
		\notag\\
	&+ D (\bm M_{\ell+1} - \bm M_{\ell-1}) \times \hat {\bm z}
		+H \hat {\bm x} 
	\label{eq:mf_heff}
\end{align}
and the constant term $C_{\ell} = ({\bm H}^{\rm eff}_{\ell} - H\hat{\bm x}) \cdot {\bm M}_{\ell}/2$. Here $N_{2{\rm d}}$ is the number of sites in each layer. ${\bm M}_{\ell}$ denotes the thermal average of spin ${\bm S}_{\bm{i}}$ in the $\ell$-th layer
\begin{equation}
{\bm M}_{\ell}=\langle \frac{1}{N_{2{\rm d}}}\sum_{\bm{i}_\bot}{\bm S}_{\bm{i}_\bot+\ell\hat{\bm{z}}}\rangle=S f(\beta S|{\bm H}^{\rm eff}_{\ell}|)
	\frac{{\bm H}^{\rm eff}_{\ell}}{|{\bm H}^{\rm eff}_{\ell}|},
	\label{eq:mf_M}
\end{equation}
with 
\begin{equation}
f(x)=\coth(x)-\frac1x. 
\end{equation}

We solve the MF equation (\ref{eq:mf_heff}),(\ref{eq:mf_M}) self-consistently by using the iterative method under the periodic boundary condition for a three-dimensional lattice with $N=N_{\rm 2d}N_z$ sites, where $N_z$ is the number of the layers.
We prepare many initial spin configurations in the form
\begin{align}
	\bm{M}_{\ell }= S (\cos k\ell, \sin k\ell, 0)
	\label{eq:theta_w}
\end{align}
with $k = 2\pi w/N_z$ for integer $w$. After solving the MF equations for each initial state, we pick up the final state that minimizes the free-energy among the configurations generated by these initial states. 
The free energy is given by the following form in the MF approximation: 
\begin{align}
	\frac{F}{N}
	= -\frac{1}{\beta N_z} \sum_{\ell} \log 
		\left(
			\frac{\sinh(\beta S|{\bm H}^{\rm eff}_{\ell}| )}{\beta S |{\bm H}^{\rm eff}_{\ell}|}
		\right)
	+\frac{1}{N_z}\sum_{\ell} C_{\ell}. 
	\label{eq:F_MF}
\end{align}
See Ref.~\cite{Shinozaki} for more details of the algorithm.


\section{\label{sec:Phase diagram}Phase diagram}
In this section, we first summarize our results on the MF phase diagram. We then discuss the underlying  physics in terms of various crossover lines. These results are helpful to understand the energetics that stabilizes a fan-type structure in chiral magnets.
\subsection{\label{subsec:Phase boundary}Phase boundary and thermodynamic phases}
\begin{figure}[tb]
	\begin{center}
		\includegraphics[width=8.5cm,clip]{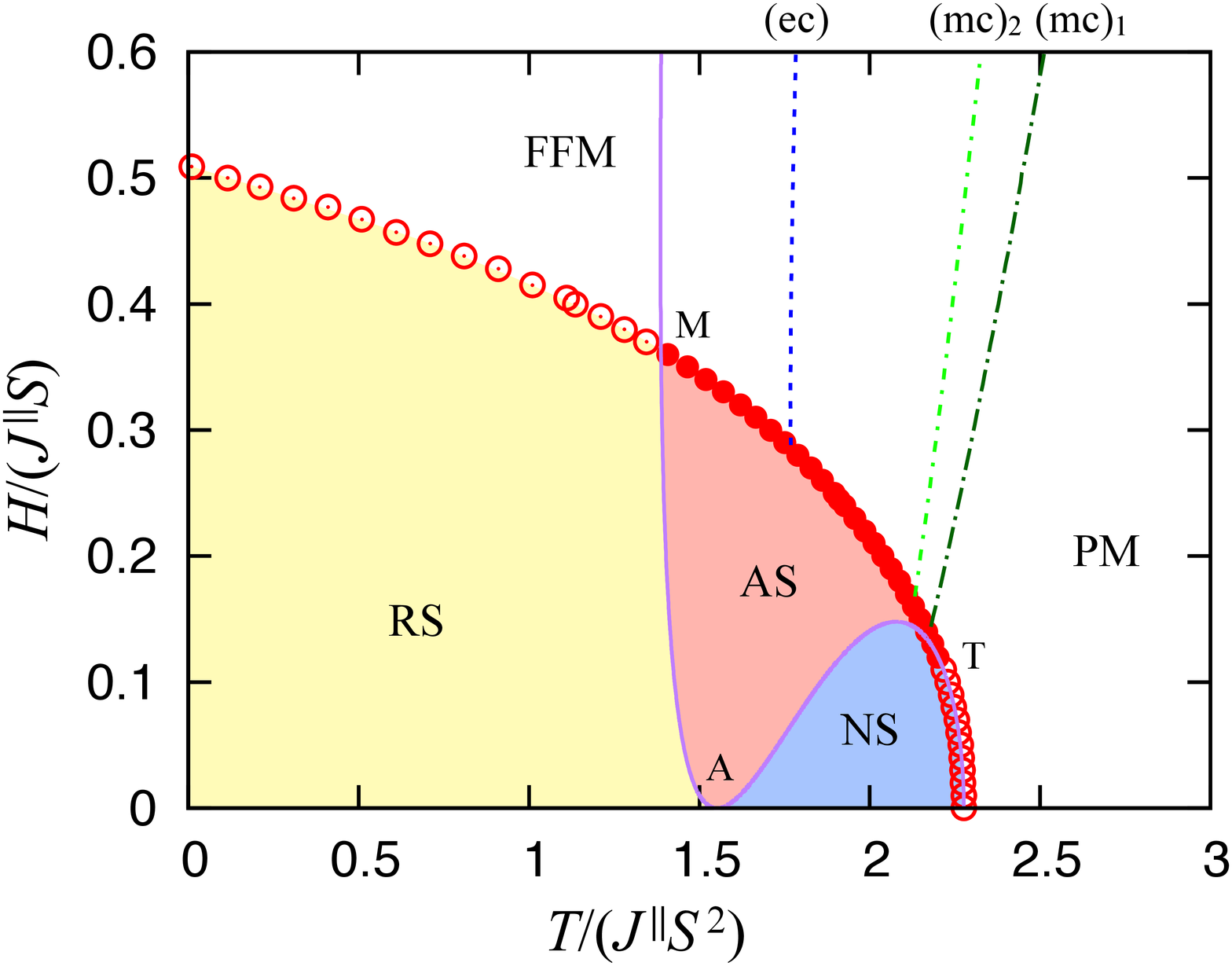}
	\end{center}
	\caption{(Color online) Phase diagram for $J^\bot/J^\parallel=1$, $D/J^\parallel =1$. The open and solid circles are, respectively, the second- and first-order phase transition points from the chiral soliton lattice (CSL) phase. 
	T and M respectively denote a tri-critical point $(T_{\rm tri}/(J^\parallel S^2), H_{\rm tri}/(J^\parallel S))\simeq (2.210,0.115)$ and a multi-critical point $(T_{\rm multi}/(J^\parallel S^2), H_{\rm multi}/(J^\parallel S))\simeq (1.396, 0.362)$. RS (AS) denotes the region where the solution of repulsively (attractively) interacting soliton exists and NS the region where no solution containing any types of solitons. The point A, which is on the horizontal axis, denotes the meeting point of the RS-AS boundary and AS-NS boundary. 
	}
	\label{fig: phase-diagram}
\end{figure}
Figure~\ref{fig: phase-diagram} shows a phase diagram in the $H-T$ plane for $J^\bot/J^\parallel =1$ and $D/J^\parallel =1$.
The first-order (discontinuous) phase transition occurs at each red solid circle.  The red open circles located at lower temperature side of the multi-critical point M represent the boundary of the continuous phase transition of nucleation-type. 
The red open circles located at higher temperature side of the tri-critical point T represent the boundary of the continuous phase transition of instability-type. The phase boundary is determined from the discontinuity (for first-order phase transition) or cusp (for continuous transitions) of the magnetization curves. 

There are only two thermodynamic phases; at the lower temperature side of the phase boundary, the spin structure exhibits one-dimensional periodic modulation with the propagation direction parallel to the $z$-axis. We call this phase chiral soliton lattice adhering to the earlier literatures~\cite{Kishine_Ovchinnikov,Togawa_symmetry}. This term implies that the modulation structure consists of underlying isolated chiral soliton ($2\pi$ domain wall). As a due caution, we note that the solution of the isolated soliton to the MF equation does not necessarily exist in the whole region of CSL phase (see III C.). At the higher temperature side of the phase boundary, the system is in the spatially uniform phase. 

We have the same structure of the phase diagram  for more realistic parameters. 
For $D/J^\parallel = 0.16$ and $J^\bot/J^\parallel =8$ corresponding to \Cr, the tri-critical point exists at $(T_{\rm tri}/(J^\parallel S^2), H_{\rm tri}/(J^\parallel S)) \simeq (11.34 , 0.00027)$ and the multi-critical point exists at $(T_{\rm multi}/(J^\parallel S^2), H_{\rm multi}/(J^\parallel S)) \simeq (11.308, 0.0011) $, where the transition temperature at zero field is $T_{\rm c}(H=0)/(J^\parallel S^2)=11.34181$ and the critical field at zero temperature is $H_{\rm c}(T=0)/(J^\parallel S) = 0.0157 $.
For $D/J^\parallel \to \infty$ with $J^\bot/D=1$, which is an extreme case, we confirm that the multi-critical point goes to $T_{\rm multi} = 0$. The tri-critical point is $ (T_{\rm tri}/(DS^2), H_{\rm tri}/(DS)) \simeq (1.7,0.58)$ where the critical field at zero temperature is $H_{\rm c}(T=0)/(DS) = 1.23687$ and the transition temperature at zero field is $T_{\rm c}(H=0)/(DS^2) = 2$.  
The structure (properties of phase transition and multi-critical points) of the phase diagram mentioned above is different from those in Refs.~\cite{Laliena2,Nishikawa} but it is essentially the same as that in Ref.~\cite{Schaub} for a continuum Ginzburg-Landau model.

\subsection{\label{subsec:Crossover lines uniform phase}
Crossover lines in spatially uniform phase}
The low temperature and high field region in the spatially uniform phase is often called ^^ ^^ forced ferromagnetic state (FFM)" and high temperature and low field region is called paramagnetic state (PM), in order to emphasize quantitative difference between the two regions. FFM has large magnetization and lower entropy and PM has small magnetization and large entropy. There is no sharp boundary between FFM and PM. However, we draw two kinds of lines as a guide of the crossover between the two. 

We draw the dashed line obtained from the peak position of specific heat as a function of temperature. We call this line ^^ ^^ entropic crossover" (ec) in the sense that entropy is released when we cross this line from the higher temperature side to the lower temperature side. Focusing on the magnetic property, we draw dotted-dashed line (we call ^^ ^^ magnetic crossover" (mc)$_1$). This line is determined from the peak position of the uniform susceptibility as a function of temperature. The magnetization starts to grow when we cross this line from high temperature to low temperature sides. In addition, we draw another line (mc)$_2$ from the condition that the second derivative of the magnetization with respect of the magnetic field becomes maximum as a function of temperature. On the basis of our result, we call the low temperature side of entropic crossover line FFM and high temperature side PM. At the transient region located between the entropic crossover and magnetic crossover lines, the system has large magnetization and large entropy.  

We note that the magnetic crossover line meets the phase boundary near the tri-critical point T. This implies that the magnetic crossover is more relevant to the property of the phase transition than the entropic crossover. 
We infer that the magnetic crossover line (mc)$_1$ is related to that observed experimentally, for example, in the uni-axial chiral helimagnet \Cr~\cite{Tsuruta2}. 
\begin{figure}[tb]
	\begin{center}
		\includegraphics[width=7cm,clip]{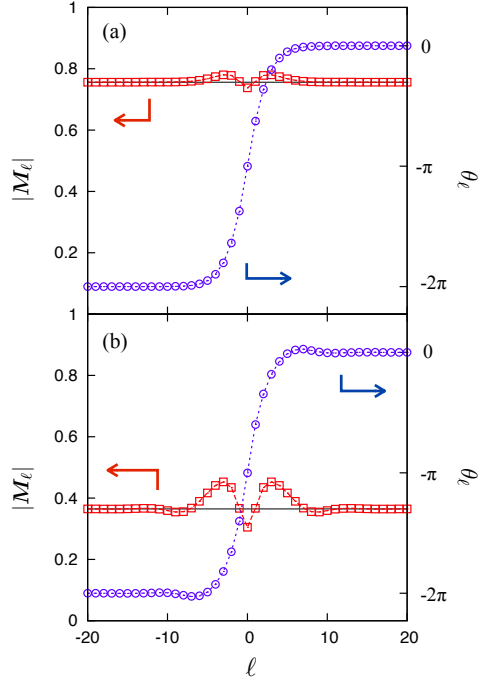}
	\end{center}
	\caption{(Color online) Spin configurations at  (a) $(T/(J^\parallel S^2), H/(J^\parallel S)) = (1.2, 0.375)$ (repulsive) and 
	(b)$(T/(J^\parallel S^2), H/(J^\parallel S)) = (2.0, 0.2)$ (attractive) with $D/J^\parallel =1$, $J^\bot/J^\parallel =1$. 
	Here, $\theta_{\ell}$ is the angle of $\bm{M}_\ell$ at $\ell$-th layer measured from the $x$-axis. 
	The black solid lines represent the asymptotic values of $\bm{M}_\ell$.}
	\label{fig: soliton}
\end{figure}
\subsection{\label{subsec:Crossover lines modulation phase}
Crossover lines in modulation phase}

As we remarked in the previous section, it is nontrivial whether the ordered phase can be always called ^^ ^^ chiral soliton lattice";  In fact we will see that no solutions containing isolated solitons exist in the region with high temperature and low field of the modulated phase. In this section, we separate the ordered phase into three regions according to the following criteria: 
\begin{itemize}
\item
the existence/absence of solutions with an isolated soliton
\item
repulsive/attractive interaction between a pair of isolated solitons.
\end{itemize}
This kind of analysis has been done for a Ginzburg-Landau-type model in Refs.~\cite{Schaub,Yamashita-Tamada,YamashitaJPSJ} and for cubic chiral helimagnet in Refs.~\cite{Leonov2010, Wilhelm1}. 
We define the isolated soliton (which has been called single discommensuration in Ref.~\cite{Schaub}) in the lattice model (\ref{eq:hamiltonian}) as the solution satisfying
\begin{equation}
\lim_{\ell \rightarrow \pm \infty} |\bm{M}_{\ell}|=M_{\rm c} 
\end{equation} 
and
\begin{equation}
\theta_{\ell}=\left\{
\begin{array}{rl}
-2\pi\quad&,\ell\rightarrow -\infty\\
-\pi\quad&,\ell=0\\
0\quad&,\ell\rightarrow \infty\\
\end{array}\right.
\end{equation}
Here $\theta_{\ell}$ is the angle between $\bm{M}_{\ell}$ and the $x$-axis, where $\bm{M}_\ell =|\bm{M}_\ell|(\cos\theta_\ell,\sin\theta_\ell,0)$ satisfying
\begin{equation}
|\theta_\ell- \theta_{\ell+1}|< \pi. 
\end{equation}
Typical profiles of isolated solitons are shown in Fig.~\ref{fig: soliton}. We see that both $|\bm{M}_\ell|$ and $\theta_\ell$ change monotonically with increasing $\ell (\ge 0)$ in Fig.~\ref{fig: soliton}(a) while they exhibit oscillating behaviors in Fig.~\ref{fig: soliton}(b). 
The modulation phase has the solutions containing an isolated soliton only at the lower temperature side of A-T line in Fig.~\ref{fig: phase-diagram}. The line A-M in Fig.~\ref{fig: phase-diagram} separates the region where an isolated soliton has a non-oscillating profile and that where an isolated soliton has $|\bm{M}_\ell|$ and $\theta_\ell$ oscillating as functions of $\ell(\ge 0)$. The details of analysis to derive the crossover lines (A-T and A-M) will be referred to the appendix. The earlier works in a GL theory~\cite{Schaub,Yamashita-Tamada, YamashitaJPSJ} and cubic chiral helimagnets~\cite{Leonov2010,Wilhelm1} have found that isolated soliton interacts repulsively (attractively) with another soliton when each isolated soliton has non-oscillating (oscillating) $|\bm{M}_\ell|$ and $\theta_\ell$. We thus separate the ordered phase into three regions: 
\begin{itemize}
\item RS: the low temperature side of the A-M line where repulsive solitons exist
\item AS: region between the A-M and the A-T lines, where attractive solitons exist
\item NS: high temperature side of the A-T line, where no solutions with any types of isolated solitons exist.   
\end{itemize}
The region ^^ ^^ NS" corresponds to the ^^ ^^ confinement region" in the cubic chiral helimagnets in Refs.~\cite{Leonov2010,Wilhelm1}. 

We confirm with our numerical accuracy that the multicritical point M determined by the thermodynamic quantities is located on the boundary between RS and AS and the tricritical point T is ^^ ^^ the triple point", where AS, NS, and the uniform phase meet. This observation implies that the property of the underlying isolated soliton governs the type of the phase transition between the modulation phase and uniform phase in the sense that
\begin{itemize}
\item The phase transition between RS and uniform phase (FFM) is nucleation-type continuous transition.
\item The phase transition between AS and uniform phase (FFM-PM) is nucleation-type discontinuous transition.
\item The phase transition between NS and uniform phase (PM) is instability-type continuous transition.   
\end{itemize}
We consider that the RS and AS can be called CSL but we expect that the NS region exhibits a different property from CSL. In the next chapter, we show that the spin structure different from CSL emerges near the phase boundary of the instability-type continuous phase transition.

\section{\label{sec:Chiral-fan}Chiral-fan structure}
\begin{figure}[tb]
	\begin{center}
		\includegraphics[width=9cm,clip]{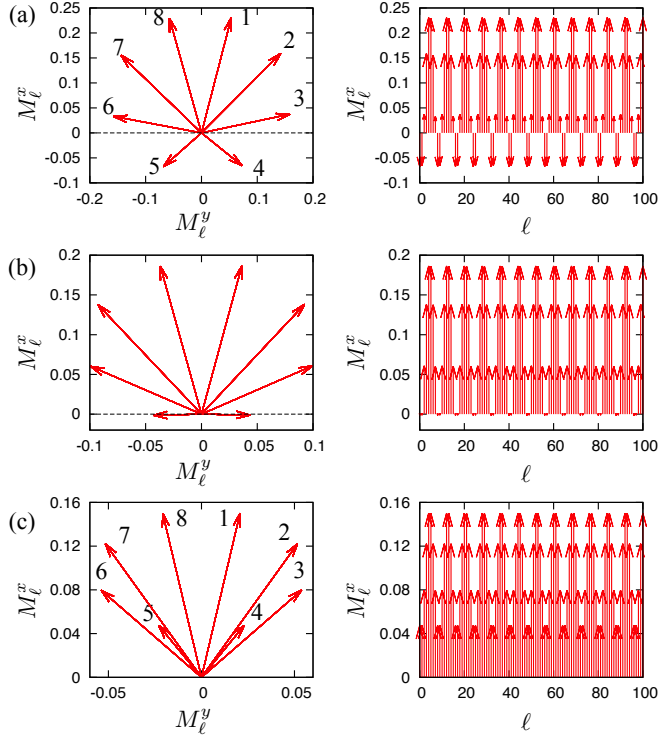}
	\end{center}
	\caption{(Color online) Spin structures with $D/J^\parallel=1, J^\bot/J^\parallel = 1$, $H/(J^\parallel S) =0.08$ for several temperatures (a) $T/(J^\parallel S^2) =$ 2.22, (b) 2.2375, (c) 2.245, where $T_{\rm c}(H)/(J^\parallel S^2) = 2.276$. Figures in the left column are the view of the spin structure from the $z$-axis. Those in the right column show the layer-dependence of the $x$-component of $\bm{M}_\ell$. 
	}
	\label{Fan_D1}
\end{figure}
\begin{figure}[tb]
	\begin{center}
		\includegraphics[width=9cm,clip]{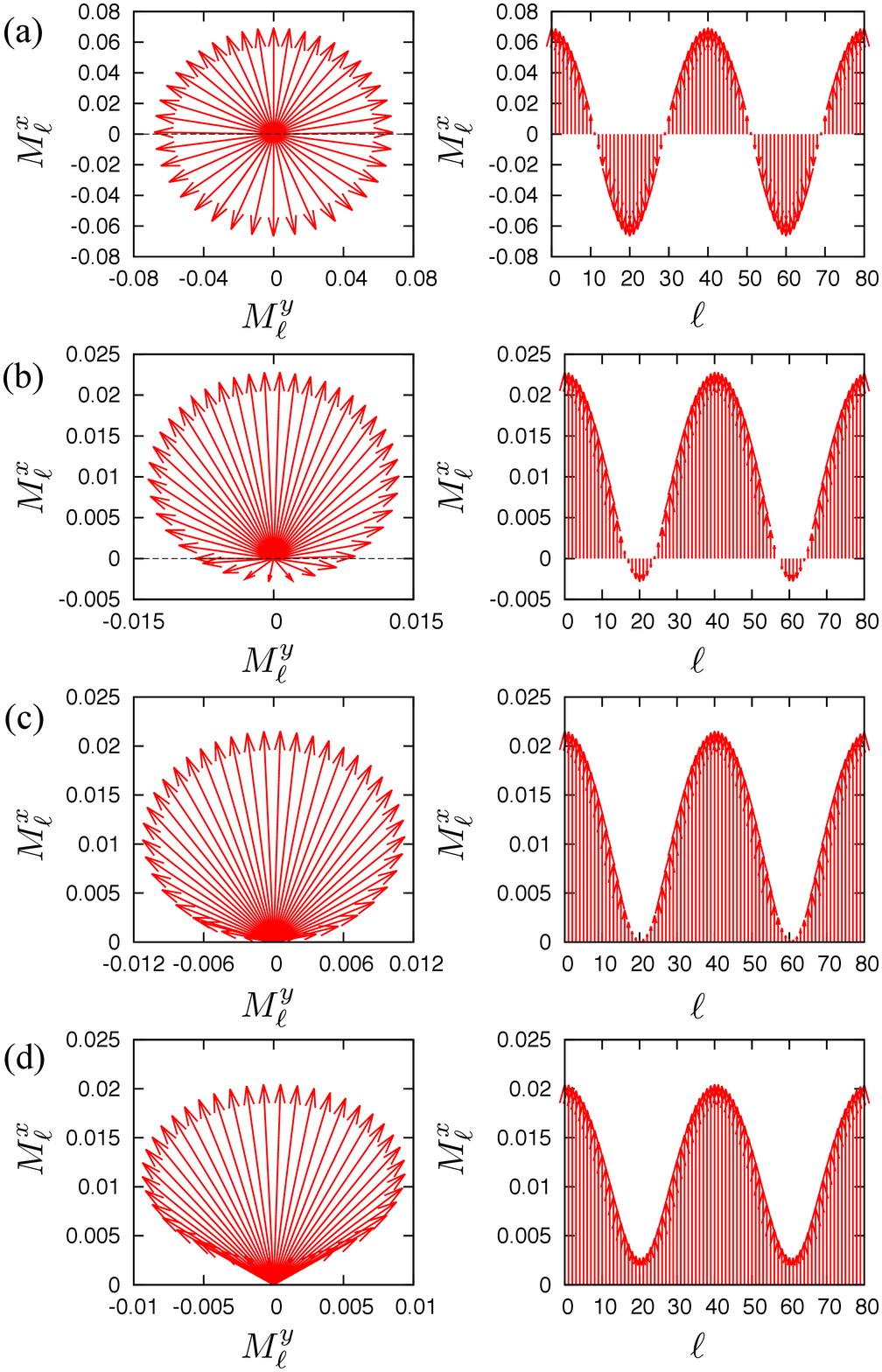}
	\end{center}
	\caption{(Color online) Spin structures with $D/J^\parallel=0.16, J^\bot/J^\parallel = 8$ (which are the parameters of \Cr), $H/(J^\parallel S) =0.0003$ for several temperatures (a) $T/(J^\parallel S^2) =$ 11.31, (b) 11.3391, (c) 11.33925, (d) 11.339315, where $T_{\rm c}(H)/(J^\parallel S^2) = 11.339317$. }
	\label{Fan_D016}
\end{figure}

Figures~\ref{Fan_D1}(a)-(c) show magnetic evolution of $\bm{M}_\ell$ with increasing $\ell$ at the parameters $D/J^\parallel=1$, $J^\bot/J^\parallel=1$, and $H/(J^\parallel S) = 0.08$.
Each red arrow represents $\bm{M}_\ell$ and the numbers in the figure represent the layer indices $\ell$.
In Figs.~\ref{Fan_D1}(a) and (b), $\bm{M}_\ell$ rotates clockwise with  increasing $\ell$ so that the DM interaction in the Hamiltonian takes negative value
$-D\hat{\bm{z}}\cdot (\bm{M}_\ell \times \bm{M}_{\ell+1})<0$.
 
In Fig.~\ref{Fan_D1}(c), however, $\bm{M}_\ell$ rotates counter-clockwise from $\ell=3$ to 6 with the modulus smaller than $\ell=1,2,7,8$. 
This sense of rotation is energetically unfavorable for the DM interaction but favorable to the Zeeman energy. 

\begin{figure}[tb]
	\begin{center}
		\includegraphics[width=7cm,clip]{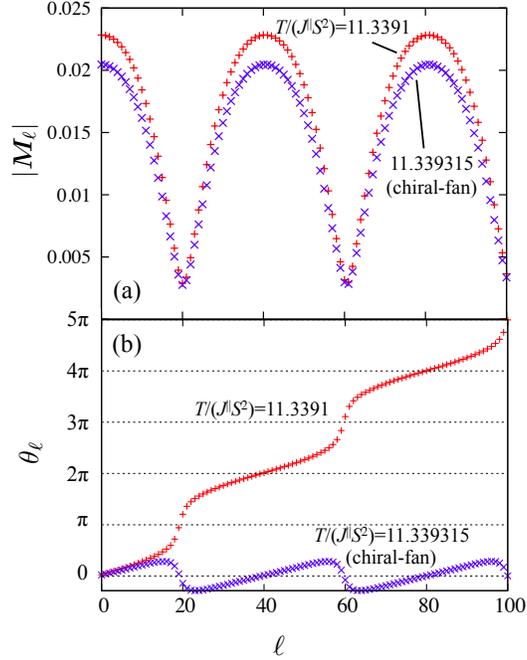}
	\end{center}
	\caption{(Color online) (a) Modulus of $\bm{M}_\ell$ and (b) angle $\theta_\ell$ measured from the $x$-axis for the chiral soliton lattice (CSL) and chiral-fan (CF) structures at $H/(J^\parallel S) =0.0003$ with $D/J^\parallel=0.16$ and $J^\bot/J^\parallel = 8$ showing  }
	\label{chiral_fan_Mandtheta}
\end{figure}

\begin{figure}[tb]
	\begin{center}
		\includegraphics[width=6cm,clip]{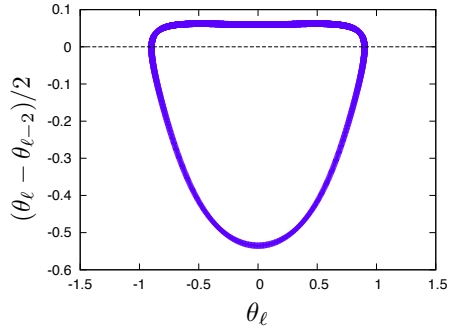}
	\end{center}
	\caption{(Color online) Phase plane trajectory of the chiral-fan structure with $D/J^\parallel=0.16, J^\bot/J^\parallel = 8$ at $H/(J^\parallel S) =0.0003$, $T/(J^\parallel S^2) =$ 11.339315.}
	\label{chiral_fan_dtheta}
\end{figure}

Figures~\ref{Fan_D016}(a)-(d) show $\bm{M}_\ell$ for $D/J^\parallel=0.16$ and $J^\bot/J^\parallel=8$, which corresponds to \Cr. 
The behaviors are similar to those in Fig.~\ref{Fan_D1}; $\bm{M}_\ell$ rotates clockwise with increasing $\ell$ in Figs.~\ref{Fan_D016}(a) and (b) while it rotates clockwise or counterclockwise alternately in Fig.~\ref{Fan_D016}(d). 
In Fig.~\ref{chiral_fan_Mandtheta}, we compare the modulus of $\bm{M}_\ell$ and the angle $\theta_\ell$  for Fig.~\ref{Fan_D016}(b) with those for Fig.~\ref{Fan_D016}(d). While the angle $\theta_\ell$ monotonically increases $2\pi$ for a period for Fig.~\ref{Fan_D016}(b), it does not grow for Fig.~\ref{Fan_D016}(d). 
Figure~\ref{chiral_fan_dtheta} shows the phase trajectory (phase portrait in the terminology of differential equation) for Fig.~\ref{Fan_D016}(d). 

In Fig.~\ref{Fan_D1}(c) and Fig.~\ref{Fan_D016}(d), we see that 
\begin{enumerate}
\item The angle $\theta_\ell$ is confined to the interval $|\theta_\ell|<\theta_{\rm c}$ with an angle $0<\theta_{\rm c}<\pi/2$.
\item $\bm{M}_\ell$ rotates clockwise or counterclockwise alternately.
\item The modulus of $\bm{M}_\ell$ becomes small during the counterclockwise rotation. 
\end{enumerate}
Obviously, the chiral soliton lattice does not have the first two properties but they are typical properties of fan structure in non-chiral (i.e. Yoshimori-type) helimagnet. The property 3 yields a difference from the conventional fan structure. We thus call the structure with the above three properties ^^ ^^ chiral-fan structure" in the present paper. 

This chiral-fan structure can be stabilized when the energy gain of the Zeeman effect dominates over the energy cost due to the DM interaction energy
\begin{align}
	D|{\bm M}_{\ell}||{\bm M}_{\ell +1}| < H |{\bm M}_{\ell}| 
\label{eq: condition}
\end{align}
in the counterclockwise rotation of ${\bm M}_{\ell}$.
This condition is satisfied in the vicinity of the transition temperature, where $|{\bm M}_{\ell}|$ becomes small. 
The reduction of the modulus in the counterclockwise rotation (listed above as the property 3) can be also understood similarly in terms of this energetics of competing two energies.

\begin{figure}[tb]
	\begin{center}
		\includegraphics[width=8cm,clip]{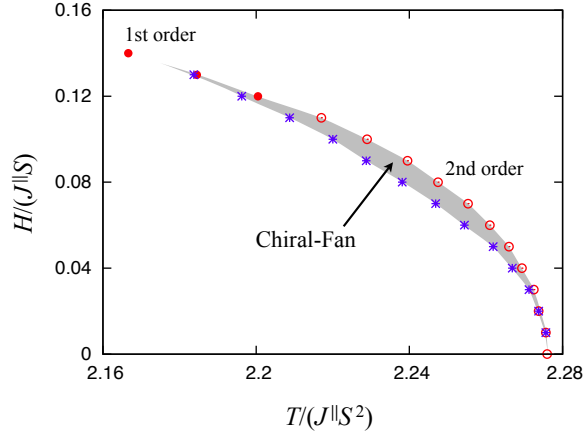}
	\end{center}
	\caption{(Color online) The magnified picture of the phase diagram around the transition temperature at low field. The dashed area between the blue cross symbols and phase boundary is the chiral-fan region. Parameters are the same as those in Fig.~\ref{fig: phase-diagram}.}
	\label{chiral_fan_region}
\end{figure}
The chiral-fan structure appears in a very small region which is indicated in Fig.~\ref{chiral_fan_region} as the colored region sandwiched between the blue cross symbols and phase boundary (red solid circles). 
\section{Conclusion\label{sec:Conclusion}}

We found a fan-type modulation structure, which we call chiral-fan structure,  near the transition temperatures in low fields perpendicular to the propagating direction of the modulation in the uni-axial chiral magnet within the MF theory. In order to examine an underlying energetics of the fan-type structure, we reconsidered the phase diagram in the whole range of the magnetic field-temperature plane and added several crossover lines in the spatially uniform phase (disordered phase) and the modulation phase (ordered phase).  

The chiral-fan structure appears near the phase boundary of the instability-type continuous transition at high temperatures and low fields. The solution of the chiral-fan exists only when the MF equation has no solutions containing any type of the isolated solitons. 

The emergence of the chiral-fan structure, the instability-type continuous transition and absence of the isolated soliton are commonly attributed to the reduction of the modulus of the local magnetization near the transition temperature in low fields.

\begin{acknowledgments}
The authors thank M.~Mito, H.~Ohsumi and Y.~Togawa for informative discussions and K. Inoue for his continuous encouragement. The computations in this work has been done using the facilities of the Supercomputer Center, the Institute for Solid State Physics, the University of Tokyo. This work was supported by JSPS KAKENHI Grant Number JP16J00091, JP17H02923, and JP25220803. This work was also supported by the Center for Chiral Science in Hiroshima University, JSPS and RFBR under the Japan - Russia Research Cooperative Program, and JSPS Core-to-Core Program, A. Advanced Research Networks. 

\end{acknowledgments}

\appendix
\section{Linear analysis of asymptotic behavior of isolated soliton}
We consider an isolated soliton to be centered at $\ell=0$. We seek for the solution to the MF equations (\ref{eq:mf_heff}) and (\ref{eq:mf_M}) and make the ansatz
\begin{subequations}
\begin{align}
&|\bm{M}_\ell|=\sqrt{(M_\ell^x)^2 +(M_\ell^y)^2}\sim M_{\rm c}+A_1 {\rm e}^{-\kappa \ell}, 
\label{eq:01}\\
&\theta_\ell \sim A_2 {\rm e}^{-\kappa \ell},
\label{eq:02}
\end{align}
\end{subequations}
for $\ell \gg 1$ with $A_1$ and $A_2$ being small. Here the modulus of the moment in the disordered phase (the spatially uniform phase) $M_{\rm c}$ is given by the solution of the equation
\begin{equation}
M_{\rm c}=S f(\beta S H_{\rm c}),\quad H_{\rm c}=(4J^{\perp}+ 2J^{\parallel})M_{\rm c}+H.\label{eq: McHc}
\end{equation}
Equations (\ref{eq:01}) and (\ref{eq:02}) are equivalent to 
\begin{equation}
\bm{M}_\ell \sim \left(M_{\rm c}+A_1 {\rm e}^{-\kappa \ell}, M_{\rm c}A_2 {\rm e}^{-\kappa \ell},0\right)\label{eq: MxMy}
\end{equation}
up to the linear order of $A_1$ and $A_2$. Substituting (\ref{eq: MxMy}) into eq.~(\ref{eq:mf_heff}) and using the second equation of (\ref{eq: McHc}), we obtain 
\begin{subequations}
\begin{align}
&H_\ell^{{\rm eff},x}=H_{\rm c}+((4J^{\perp} +2J^{\parallel}\cosh \kappa) A_1 -2D A_2 M_{\rm c}\sinh \kappa)e^{-\kappa \ell}
\label{eq: Mx}\\
&H_\ell^{{\rm eff},y}=\left((4J^{\perp} +2J^{\parallel} \cosh \kappa) A_2 M_{\rm c}+2D A_1\sinh \kappa\right)e^{-\kappa \ell}.
\label{eq: My}
\end{align}
\end{subequations}
From Eqs.~(\ref{eq:mf_M}) and (\ref{eq: MxMy}), it follows that 
\begin{subequations}
\begin{align}
&A_1 e^{-\kappa \ell}=f'(\beta S H_{\rm c}) \beta S^2 (H_\ell^{{\rm eff},x}-H_{\rm c}),
\label{eq: Hx-linear}\\
&M_{\rm c}A_2 e^{-\kappa \ell}=\frac{M_{\rm c}H_\ell^{{\rm eff},y}}{H_{\rm c}}.
\label{eq: Hy-linear}
\end{align}
\end{subequations}
Eliminating $H_\ell^{{\rm eff},x}$ from (\ref{eq: Mx}), (\ref{eq: My}) and $H_\ell^{{\rm eff},y}$  from (\ref{eq: Hx-linear}), (\ref{eq: Hy-linear}), we obtain %
\begin{equation}
\begin{pmatrix}
(4J^{\perp} +2J^{\parallel} \cosh \kappa)\beta S^2 f'(\beta S H_{\rm c})-1 & -2D M_{\rm c} \beta S^2 f'(\beta S H_{\rm c})\sinh\kappa\\
2D\sinh\kappa & (4J^{\perp} +2J^{\parallel} \cosh \kappa)M_{\rm c}-H_{\rm c}\\
\end{pmatrix}
\begin{pmatrix}
A_1\\
A_2\\
\end{pmatrix}
=0. \label{eq: A1A2}
\end{equation}
A nontrivial solution to (\ref{eq: A1A2}) exists when 
\begin{equation}
\mathcal{A} \cosh^2\kappa +\mathcal{B}\cosh\kappa +\mathcal{C}=0
\label{eq: quadratic}
\end{equation}
with 
\begin{subequations}
\begin{align}
&\mathcal{A}=4 M_{\rm c}\beta S^2 ((J^{\parallel})^2 +D^2)f'(\beta S H_{\rm c})
\label{eq: mathcalA}\\
&\mathcal{B}=2J^{\parallel}  \left(
(8\beta S^2 J^{\perp} M_{\rm c}f'(\beta S H_{\rm c})-M_{\rm c}-
\beta S^2 H_{\rm c}f'(\beta S H_{\rm c})\right)
\label{eq: mathcalB}\\
&\mathcal{C}=(4\beta S^2 J^{\perp} f'(\beta S H_{\rm c})-1)(4J^{\perp} M_{\rm c}-H_{\rm c})-4 M_{\rm c}\beta S^2 D^2 f'(\beta S H_{\rm c}).
\label{eq: mathcalC}
\end{align}
\end{subequations}
From (\ref{eq: quadratic}), we obtain the three cases:
\begin{itemize}
\item real $\kappa$: discriminant $\mathcal{B}^2 -4 \mathcal{A}\mathcal{C}>0$
and  $\cosh\kappa=\frac{-\mathcal{B}\pm \sqrt{\mathcal{B}^2 -4 \mathcal{A}\mathcal{C}}}{2\mathcal{A}}>1$
\item pure imaginary $\kappa$: discriminant $\mathcal{B}^2 -4 \mathcal{A}\mathcal{C}>0$ and $-1<\frac{-\mathcal{B}\pm\sqrt{\mathcal{B}^2 -4 \mathcal{A}\mathcal{C}}}{2\mathcal{A}}<1$
\item complex $\kappa$: discriminant $\mathcal{B}^2 -4 \mathcal{A}\mathcal{C}<0$,
\end{itemize}
from which we can draw the crossover lines shown in Fig.~\ref{fig: phase-diagram}. 
%
\newpage 
\bibliography{apssamp}

\begin{thebibliography}{99}
%
\bibitem{Izyumov} Yu.~A.~Izyumov, Sov. Phys. Usp. {\bf 27}, 845 (1985). 
\bibitem{Yoshimori} A.~Yoshimori, 
	{\it J. Phys. Soc. Jpn.} {\bf 14}, 807 (1959).
\bibitem{Dzyaloshinskii1} I.~E.~Dzyaloshinski, 
	{\it Soviet Physics JETP} {\bf 19}, 960 (1964).
\bibitem{Dzyaloshinskii2} I.~E.~Dzyaloshinskii, 
	{\it Soviet Physics JETP} {\bf 20}, 223 (1965). 
\bibitem{Dzyaloshinskii3} I.~E.~Dzyaloshinskii, 
	{\it Soviet Physics JETP} {\bf 20}, 665 (1965). 
\bibitem{Bogdanov89}
A.~N.~Bogdanov and D.~A.~Yablonskii, 
        {\it Soviet Physics JETP} {\bf 68}, 101 (1989).
\bibitem{Bogdanov94}
A.~Bogdanov and A.~Hubert, {\it J. Magn. Magn. Mater.} {\bf 138}, 255 (1994); {\bf 195}, 182 (1999).
%
\bibitem{Kishine_S_T} J.~Kishine, K.~Inoue, and Y.~Yoshida, 
	{\it Progress of Theoretical Physics Supplement} {\bf 159}, 82 (2005).
\bibitem{Moriya} T.~Moriya and T.~Miyadai,
	{\it Solid State Communications} {\bf 42}, 209 (1982).
\bibitem{Miyadai} T.~Miyadai, K.~Kikuchi, H.~Kondo, S.~Sakka, M.~Arai, and Y.~Ishikawa, 
	{\it J. Phys. Soc. Jpn.} {\bf 52}, 1394 (1983).
\bibitem{Togawa1} Y.~Togawa, T.~Koyama, K.~Takayanagi, S.~Mori, Y.~Kousaka, J.~Akimitsu, S.~Nishihara, K.~Inoue, A.~S.~Ovchinnikov, and J.~Kishine, 
	{\it Physical Review Letters} {\bf 108}, 107202 (2012). 
\bibitem{Kishine_Ovchinnikov} J.~Kishine and A.~S.~Ovchinnikov,
	{\it Solid State Physics} {\bf 66}, 1 (2015).
\bibitem{Togawa_symmetry} Y.~Togawa, Y.~Kousaka, K.~Inoue and J.~Kishine, 
	{\it J. Phys. Soc. Jpn.} {\bf 85}, 112001 (2016).
\bibitem{deGennes}P. de Gennes, in Fluctuations, Instabilities and Phase transitions, Ed. T. Riste, NATO ASI Series B Vol. 2(Plenum, New York, 1975)
\bibitem{Bulaevskii}L.~N.~Bulaevslii and V.~L.~Ginzburg, 
	{\it Soviet Physics JETP} {\bf 18}, 530 (1964). 
\bibitem{Schaub} B.~Schaub and D.~Mukamel, 
	{\it Physical Review B} {\bf 32}, 6385 (1985).
\bibitem{Leonov2010}A.~A.~Leonov, A.~N.~Bogdanov and U.~K.~R\"o\ss ler, arXiv: 1001.1292v3 (2010). 
\bibitem{Wilhelm1} H.~Wilhelm, M.~Baenitz, M.~Schmidt, C.~Naylor, R.~Lortz, U.~K.~R\"{o}\ss ler, A.~A.~Leonov, and A.~N.~Bogdanov, 
	{\it Journal of Physics: Condensed Matter} {\bf 24}, 294204 (2012).
\bibitem{Tsuruta2} K.~Tsuruta, M.~Mito, H.~Deguchi, J.~Kishine, Y.~Kousaka, J.~Akimitsu, and K.~Inoue,
	{\it Physical Review B} {\bf 93}, 104402 (2016).

\bibitem{Laliena1} V.~Laliena, J.~Campo, J.~Kishine, A.~S.~Ovchinnikov, Y.~Togawa, Y.~Kousaka, and K.~Inoue, 
	{\it Physical Review B} {\bf 93}, 134424 (2016).
\bibitem{Laliena2} V.~Laliena, J.~Campo, and Y.~Kousaka,
	{\it Physical Review B} {\bf 94}, 094439 (2016).
\bibitem{Laliena3} V.~Laliena, J.~Campo, and Y.~Kousaka, 
	arXiv:1610.00996v1 (2016).
	
\bibitem{Shinozaki} M.~Shinozaki, S.~Hoshino, Y.~Masaki, J.~Kishine, and Y.~Kato,
	{\it J. Phys. Soc. Jpn} {\bf 85}, 074710 (2016); 
	M.~Shinozaki, S.~Hoshino, Y.~Masaki, J.~Kishine, and Y.~Kato,
	{\it J. Phys. Soc. Jpn} {\bf 86}, 038001 (2017) (Erratum). 
	
\bibitem{Nishikawa} Y.~Nishikawa and K.~Hukushima, 
	{\it Physical Review B} {\bf 94} 064428 (2016).
\bibitem{Yamashita-Tamada} M.~Yamashita and O.~Tamada, 
	{\it J. Phys. Soc. Jpn.} {\bf 54}, 2963 (1985).
\bibitem{YamashitaJPSJ} M.~Yamashita, 
	{\it J. Phys. Soc. Jpn.} {\bf 56}, 1414 (1987).
%
%
%
\end{thebibliography}
\bibliographystyle{prsty}

\end{document}